# A Comment on "The Far Future of Exoplanet Direct Characterization" – the Case for Interstellar Space Probes

Ian A. Crawford[1]

[1]*Department of Earth and Planetary Sciences, Birkbeck College, University of London, Malet Street, London, WC1E 7HX, United Kingdom*

**Address for correspondence**: I. A. Crawford, Department of Earth and Planetary Sciences, Birkbeck College London, Malet Street, London, WC1E 7HX, United Kingdom
**Telephone**: +44 207 679 3431
**Fax**: +44 207 679 2867
**Email:** i.crawford@ucl.ac.uk

**Abstract**

Following on from ideas presented in a recent paper by Schneider et al. (2010) on "The Far Future of Exoplanet Direct Characterization," I argue that they have exaggerated the technical obstacles to performing such "direct characterization" by means of fast (order 0.1c) interstellar space probes. A brief summary of rapid interstellar spaceflight concepts that may be found in the literature is presented. I argue that the presence of interstellar dust grains, while certainly something that will need to be allowed for in interstellar vehicle design, is unlikely to be the kind of "show stopper" suggested by Schneider et al. Astrobiology as a discipline would be a major beneficiary of developing an interstellar spaceflight capability, albeit in the longer term, and I argue that astrobiologists should keep an open mind to the possibilities.

**1. Introduction**

In a recent paper, Schneider et al. (2010) considered the far future of the characterization of extrasolar planetary systems. Most of their review concerns the future of astronomical techniques for the imaging and spectroscopic analysis of such planets, and, as they argue, it seems certain that these techniques will improve considerably in sophistication and sensitivity in the coming decades. On the other hand, the history of planetary exploration in our own Solar System demonstrates that the most detailed knowledge of other planets actually results from *in situ* measurements by spacecraft. While future advances in astronomical instrumentation may be expected to fill in details of the structure and composition of exoplanet atmospheres, including the presence of biomarkers (e.g., Cockell et al., 2009; Kaltenegger et al, 2010), it seems unlikely that detailed knowledge of the surface geology, sub-surface structure, or, if present, the biology of such planets will be obtained by these techniques. One only has to look at the range of spacecraft-borne instrumentation being planned for astrobiological investigations of Mars (e.g., Parnell et al., 2007; Bada et al., 2008; Coleman and Grunthaner, 2008, and references therein) and compare them with what might be learned about present or past life on the planet by using only Earth-based astronomical instruments, to see the fundamental limitations of the latter. If biosignatures are one day detected in the atmosphere of a planet orbiting a nearby star, the limitations of astronomical remote-sensing for biological follow-up studies and the benefits of direct *in situ* investigation will only become even more apparent. For example, just how much will follow-up astronomical observations tell us about the underlying biochemistry, cellular structure, ecological diversity, and evolutionary history of any life that may be detected spectroscopically on a terrestrial planet orbiting, say, alpha Centauri B?

However, despite the wide range of scientific benefits that would result from *in situ* investigations of other planetary systems (Crawford, 2010a), interstellar space travel on timescales relevant to human society will be a formidable technological undertaking. In particular, it will require the attainment of velocities that are a significant fraction of the speed of light. Schneider et al. (2010) briefly considered this issue and concluded, probably correctly, that it may be several centuries before we are likely to attain such a capability. However, for a paper that has the words "direct characterization" in the title, I am concerned by what I see as a rather shallow and somewhat misleading treatment of the long-term possibilities for interstellar travel by Schneider et al. (2010). I therefore here briefly review some of the key issues related to the technical practicality of interstellar space travel and draw some conclusions as to its possible long-term importance for astrobiology.

**2. Interstellar travel: a brief review**

Colleagues unfamiliar with the subject may be surprised to know that there is actually quite a large literature devoted to the technical requirements of, and proposed solutions to, the problem of rapid interstellar space travel. The interested reader will find detailed reviews by Crawford (1990), Mauldin (1992), Matloff (2000), and Gilster (2004), all of whom presented many references to the primary astronautical literature. For present purposes, I define "rapid" to signify velocities of the order of ten percent of the speed of light (0.1c), which would permit travel times to the nearer stars of several decades and be sufficient for direct scientific investigation of any accompanying planetary systems.

Schneider et al. (2010) cited 0.3c as typical velocity for interstellar spaceflight proposals, but it is difficult to see how they have arrived at this value. While certainly desirable if ever achievable, 0.3c is significantly higher than appears feasible according to the more well-developed proposals in the literature, which generally yield velocities closer to 0.1c. Note that neither my earlier paper (Crawford, 1990), nor that of Bjørk (2007), both of which were cited by Schneider et al. to justify the higher value, actually advocate a velocity of 0.3c – the former paper reviews a wide range of options, while the latter study is based on the a value of 0.1c. Even the latter value will be very challenging to achieve in practice (e.g., Crawford, 1990), and 0.3c will be an order of magnitude more difficult owing the scaling of kinetic energy with the square of the velocity. By arbitrarily over-estimating the speed required, I believe that Schneider et al. may have been led to a premature rejection of the feasibility of the direct investigation of nearby stars by interstellar space probes.

The most technically mature concepts for achieving rapid interstellar travel are those based on nuclear fusion propulsion (reviewed by Cassenti, 2002), of which the *Daedalus* study (Bond et al, 1978) is still the most detailed engineering assessment available in the literature. *Daedalus* was designed to accelerate a 450 tonne scientific payload to 12% of the speed of light, using what were considered to be reasonable extrapolations of then-current technical knowledge. This would permit a travel time of 36 years to the nearest star, although the resulting vehicle would be very massive (requiring approximately 50,000 tonnes of nuclear fuel) and far beyond present capabilities to construct. That said, in the decades following the original *Daedalus* study, technical advances in a number of fields have occurred which may make fusion-powered vehicles of the *Daedalus*-type more practical as long-term solutions to the problem of interstellar travel. These include developments in miniaturization and nanotechnology, which would ensure that a much less massive payload would be required than was then assumed, and developments in inertial confinement fusion research for power generation on Earth. Indeed, the National Ignition Facility recently commissioned at the Lawrence Livermore National Laboratory in California (https://lasers.llnl.gov) is, albeit unintentionally, building up technical competencies directly relevant to the development of fusion-based space propulsion systems. For these reasons, there is a strong case for a reassessment of the *Daedalus* concept in light of updated scientific and technical knowledge, and at least one such study is currently underway (Long et al., 2010).

In addition to nuclear-based propulsion systems, other suggested methods of rapid interstellar space travel include antimatter-powered rockets (Forward, 1982; Morgan, 1982), laser-powered rockets (Jackson and Whitmire 1978), laser-pushed light sails (Forward 1984), and interstellar ramjets that would collect their fuel directly from the interstellar medium (Bussard, 1960; Whitmire, 1975). A particularly interesting version of the latter concept, which might lend itself to a relatively early implementation by removing the need to carry large quantities of nuclear fuel (as for *Daedalus*), while avoiding the significant technical challenges of collecting it from the interstellar medium (as suggested by Bussard, 1960), is the "ramjet runway" (Whitmire and Jackson, 1977; Matloff, 1979). In this concept, pellets of nuclear fuel would be deposited along the interstellar trajectory, having been launched in advance from the Solar System, for the interstellar space probe to collect and utilize as it accelerates toward its goal.

All these interstellar travel concepts, and others, are described in more detail in the reviews cited above. The diversity of these approaches, all having different strengths and weaknesses, yet all fully consistent with the laws of physics as currently understood, gives some confidence that viable solutions to the problem of interstellar space travel exist that could be utilized should a decision be made to engage in this activity. Of course, future scientific discoveries and technological improvements may make the undertaking easier than it currently appears, but it is encouraging that such a wide range of technical solutions can already be identified.

## 3. The (potential) problem of interstellar dust

Schneider et al. (2010) identified the ubiquitous presence of interstellar dust particles as the principal obstacle to rapid interstellar spaceflight. It is certainly true that impacts with interstellar grains will be potentially damaging for space vehicles travelling at a significant fraction of the speed of light, and this will need to be factored into the vehicle design. However, the problem is not as severe as implied by Schneider et al., for a number of reasons.

Firstly, as noted above, Schneider et al. have exaggerated the problem by assuming a much higher vehicle velocity (0.3c) than is likely to be achieved by even a very ambitious interstellar probe. Secondly, they have also exaggerated the size of typical interstellar grains, which in the solar neighborhood are expected mostly to be sub-micron in size (Mathis et al., 1977; Landgraf et al., 2000; Frisch et al., 2009; Draine, 2009), rather than the 100 μm adopted by Schneider et al. For example, the mass of a 1 μm-radius grain of silicate composition (i.e., a density of 2500 kg m$^{-3}$) is 10$^{-14}$ kg, and its kinetic energy at 0.1c is 4.5 J (seven orders of magnitude smaller than the value of 4×10$^7$ J we get for a 100 μm grain travelling at 0.3c as given in Schneider et al.'s example).

Until recently, it would have been expected that 1 µm would be an absolute upper-limit for the size of interstellar grains in the warm (T ~ 6000 K), low-density (n ~ 0.1 to 0.2 hydrogen nuclei $cm^{-3}$) interstellar cloud (the "Local Interstellar Cloud"; LIC) which surrounds the Sun (see Cox and Reynolds (1987) and Frisch et al. (2009) for reviews). However, recent spacecraft measurements of interstellar grains entering the outer Solar System (e.g., Landgraf et al., 2000; Frisch et al., 2009) have identified a high-mass tail to the local interstellar grain population extending to perhaps as high as $10^{-12}$ kg (i.e. 4.5 µm radius), although this value is strongly dependent on the calibration of the dust detectors on the spacecraft (compare Figures 4 and 5 of Landgraf et al., 2000). These measurements led Landgraf et al. (2000) to infer a total dust mass density in the LIC of $6.2 \times 10^{-24}$ kg $m^{-3}$. This is somewhat higher (by a factor of a few) than earlier estimates (e.g., Martin, 1978; Crawford, 1990) and would have to be allowed for in the design of an interstellar space vehicle.

The issue of shielding an interstellar space probe from interstellar dust grains was considered in detail in the context of the *Daedalus* study by Martin (1978). Martin adopted beryllium as a potential shielding material, owing to its low density and relatively high specific heat capacity, although doubtless other materials could be considered. Following Martin's (1978) analysis, but adopting an interstellar dust density of $6.2 \times 10^{-24}$ kg $m^{-3}$ (i.e., that determined by Landgraf et al., 2000), we find that erosion by interstellar dust at a velocity of 0.1c would be expected to erode of the order of 5 kg $m^{-2}$ of shielding material over a six light-year flight (see Table 6 of Martin, 1978). The need to provide such shielding will certainly add to the mass of an interstellar probe, but it hardly seems to be the "show-stopper" implied by Schneider et al (2010).

Now, it is true that the upper-bound to the size distribution of interstellar dust particles in the solar neighborhood is not well constrained (Landgraf et al., 2000; Frisch et al., 2009; Krüger and Grün, 2009). Interstellar particles even larger than the largest inferred from the spacecraft dust detectors (of the order of 40µm in size) may have been detected by meteor radar observations (e.g., Baggaley, 2000; Krüger and Grün, 2009), and the possibility of colliding with even larger particles over the course of a voyage of several light-years cannot be discounted. Indeed, converting the interstellar grain fluxes shown in Fig. 11 of Landgraf et al. (2000) to spatial densities in the LIC (following their assumption that these grains enter the Solar System with a relative velocity of 26 km s$^{-1}$), and crudely extrapolating to higher masses, we would infer the spatial density of 100-µm grains to be about $4\times10^{-17}$ m$^{-3}$. Thus, over the six light-year ($5.7\times10^{16}$ m) flight considered by Martin (1978), we might expect of the order of two impacts per square meter with such large particles, and Schneider et al.'s (2010) injunction may, after all, appear pertinent. On the other hand, it is far from clear that it is valid to extrapolate the distribution to such large masses, not least because of the difficulty of reconciling the presence of such large solid particles in the LIC with constraints imposed by the cosmic abundances of the elements [as also noted by Landgraf et al. (2000) and Draine (2009)]. Clearly, more work needs to be done to better determine the upper-limit to size distribution of interstellar dust grains in the local interstellar medium.

Nevertheless, even if future work does reveal the presence of a significant population of large interstellar grains in the LIC, it is, contrary to the impression given by Schneider et al. (2010), certainly possible to envisage appropriate countermeasures. Schneider et al. stated that "no currently available technology could protect against such a threat without a spacecraft that has a mass of hundreds of tons." However, as will be clear from the discussion in Section 2 above, *any* interstellar vehicle capable of accelerating to a significant fraction of the speed of light will have a mass of (at least) several hundred tonnes, and so the caveat added by Schneider et al. doesn't really apply. Given a sufficiently large spacecraft, one could imagine a range of sophisticated sensing instruments (e.g., based on radar or lidar) to detect large in-coming grains and employ active (e.g. laser) or passive (e.g. electromagnetic) means with which to destroy or deflect them before they impact the spacecraft. However, by far the simplest solution, developed by Bond (1978) for the *Daedalus* study, would be for the spacecraft to be preceded by a fine cloud of small dust particles (ejected from the vehicle, and thus travelling at the same velocity but a small distance ahead), such that any in-coming large grains will be destroyed by collisions within this artificial dust cloud before they have a chance to reach the main vehicle. This concept was developed for *Daedalus* in the context of protecting the vehicle in the denser interplanetary environment of a target star system, but it would work just as well for the interstellar phase of the mission should further research identify the need for such protection.

## 4. Conclusions

Journeys to the nearer stars with travel-times of decades (necessitating velocities of the order of ten percent of the speed of light) will be a considerable technological (as well as economic and political) undertaking. The magnitude of the difficulties should not be underestimated, but neither should they be exaggerated. There is a large technical literature (reviewed by Crawford, 1990; Matloff, 2000; and Gilster, 2004), which demonstrates that rapid interstellar space travel is not physically impossible and is a legitimate technological goal for the centuries ahead. Ultimately, the development of this capability may be the only way to follow-up any detections of biosignatures that may be made in the atmospheres of Earth-like planets orbiting nearby stars in the coming decades (e.g., Cockell et al., 2009; Kaltenegger et al., 2010). The history of planetary exploration in our own Solar System indicates that reliance on astronomical observations will be very limiting in this respect (as Schneider et al. themselves correctly deduce). Moreover, it may only be through such a far-reaching discovery that the collective attention of humanity is energized sufficiently to free up the intellectual, political, and financial resources that will be required if interstellar spaceflight is ever to become a reality.

Rapid interstellar space travel would yield benefits in many areas of science (especially in astrophysics and planetary science; see Crawford (2010a) for a discussion), but it seems likely that astrobiology would be the main long-term beneficiary. Thus, in spite the obvious technical hurdles, when considering "the far future of exoplanet direct characterization" (Schneider et al., 2010) astrobiologists need to keep an open mind to the possibilities of, and develop the arguments for, the long-term development of a rapid interstellar spaceflight capability. As argued elsewhere (Crawford, 2010b), this includes, in the shorter term, lending our collective weight to arguments for the development of a mature space exploration infrastructure in our own Solar System. Not only will this yield considerable astrobiological benefits closer to home (e.g., by facilitating the search for past or present life on Mars and the moons of the outer Solar System), but it is probably a prerequisite if we are to create the technological and economic conditions from which a longer-term capability of rapid interstellar spaceflight can emerge.

**Author Disclosure Statement**

No competing financial interests exist for the author of this manuscript.

**Abbreviations**

LIC, Local Interstellar Cloud